\documentclass[aps,prl,showpacs,twocolumn,superscriptaddress,citeautoscript]{revtex4}
\usepackage{graphics}
\usepackage{graphicx}
\usepackage{amsmath}
\usepackage{amsfonts}
\usepackage{epsfig}
\usepackage{comment,braket}
\usepackage{color}

\def\beq{\begin{equation}}
\def\eeq{\end{equation}}
\def\vk{{\bf k}}

\begin{document}

\title{Electronic properties and magnetism of iron at the Earth's inner core conditions}

\author{L.~V.~Pourovskii}
\affiliation{Centre de Physique Th\'eorique, CNRS, \'Ecole Polytechnique, 91128 Palaiseau, France}
\affiliation{Swedish e-science Research Centre (SeRC), Department of Physics, Chemistry and Biology (IFM), Link\"oping University, Link\"oping, Sweden}
\author{T.~Miyake}
\affiliation{Nanosystem Research Institute, AIST, Tsukuba 305-8568, Japan}
\author{S.~I.~Simak}
\affiliation{Department of Physics, Chemistry and Biology (IFM), Link\"oping University, Link\"oping, Sweden}
\author{A.~V.~Ruban}
\affiliation{Department of Materials Science and Engineering, Royal Institute of Technology, SE-10044, Stockholm, Sweden}
\author{L. Dubrovinsky}
\affiliation{Bayerisches Geoinstitut, Universit\"at Bayreuth, 95440 Bayreuth, Germany}
\author{I.~A.~Abrikosov}
\affiliation{Department of Physics, Chemistry and Biology (IFM), Link\"oping University, Link\"oping, Sweden}

\begin{abstract}
We employ state-of-the-art {\it ab initio} simulations within the dynamical mean-field theory to study three likely phases of iron (hexogonal close-packed, $hcp$, face centered cubic, $fcc$, and body centered cubic, $bcc$) at the Earth's core conditions. We demonstrate that the correction to the electronic free energy due to correlations can be significant for the relative stability of the phases. The strongest effect is observed in bcc Fe, which shows a non-Fermi liquid behaviour, and where a Curie-Weiss behaviour of the uniform susceptbility hints at a local magnetic moment still existing at 5800 K and 300 GPa. We predict that all three structures have sufficiently high magnetic susceptibility to stabilize the geodynamo.
\end{abstract}

\maketitle

Being the main component of the Earth's core, iron attracts significant attention from broad research community. Understanding its properties at ultra-high pressure and temperature, ranging from studies of the core structure to modeling the geodynamo is a long-term goal for the condensed matter physics, and is essential for explaning geochemical observations, seismic data, and for the theory of geomagnetism, to mention few examples. In spite of all previous theoretical and experimental efforts, the crystal structure and properties of solid Fe at the inner Earth core conditions remain a subject of intense debates.
All three phases stable at low pressure-temperature conditions, namely, $hcp$, $fcc$, and $bcc$, 
have been suggested as possible crystal structures  of iron or its alloys in the Earth inner core \cite{
Lin2002,Dubrovinsky2007,Belonoshko2003,Vocaldo2003,Mikhaylushkin2007}.

Theoretical simulations of iron at high pressures and temperatures generally rely on the picture of a wide-band metal 
with insignificant local correlations \cite{Belonoshko2003,Vocaldo2003,Mikhaylushkin2007,Stixrude2012}. Indeed under compression the overlap between localized states 
increases and so does the bandwidth W, while the local Coulomb repulsion U between those states is 
screened more efficiently. The reduction of the U/W ratio is used to rationalize the absence of electronic 
correlations beyond the standard local density approximation (LDA) 
 at high-pressure conditions. The increase of the $3d$-band width also results 
in the corresponding drop of the density of states at the Fermi energy leading to disappearance of 
the driving force for magnetism according to the Stoner criterion. Besides, even at the ambient pressure 
but at very high temperature T $\gg$ T$_c$ (T$_c$ is the Curie temperature, which is 1043 K in $\alpha-$Fe) local 
magnetic moments are expected to be suppressed due to one-electron Stoner-type excitations. Thus, 
when extremely high pressure and temperature are simultaneously applied, disappearance of the local 
magnetic moment seems to be inevitable. Due to these considerations iron at the Earth's inner core 
conditions has been modeled as non-magnetic within LDA-based approaches. The results of recent 
works of Sola et al. \cite{Sola2009,Sola2009_1}, who applied fixed-node-approximation 
quantum Monte Carlo techniques 
to compute the equation of state and the melting temperature of $hcp$ Fe at extreme conditions, 
are in good agreement with previous LDA-based simulations, thus strengthen the above argument, 
at least in the case of the $hcp$ phase. On the other hand, Glazyrin {\it et al.} \cite{Glazyrin2013} have just demonstrated importance of correlation effects in hcp iron revealed by an electronic topological transition induced at pressure ~40 GPa and room temperature.

\begin{figure}
\begin{center}
\includegraphics[width=0.75\columnwidth]{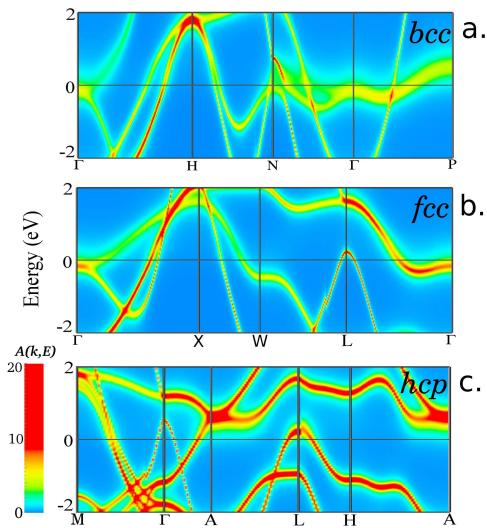}
\caption{\label{fig:band_struct} (Color online)
The LDA+DMFT k-resolved spectral function A(k,E) ($V_{at}$/eV) for $bcc$ (a), $fcc$ (b), and
$hcp$ (c) Fe at volume $V_{at}=$7.05 \AA$^3$/atom and temperature 5800~K. A non-quasiparticle $e_g$ band is seen in the vicinity of the Fermy energy along the $N-\Gamma-P$ path in (a).}
\end{center}
\end{figure}

So, are electronic correlations important at extreme conditions, and can they lead to qualitatively new phenomena? 
To address this question we have investigated the impact of correlations on the electronic structure, magnetic properties 
and thermodynamic stability of iron by 
performing {\it ab initio} simulations of the $bcc$, $fcc$ and $hcp$ phases for the volume of 7.05 \AA/atom, 
corresponding to the pressures expected in the inner Earth core,
and temperatures up to 5800~K (The $c/a$ ratio in $hcp$ Fe was fixed at 1.60 \cite{Ono2010}).
We employ a 
state-of-the-art fully self-consistent technique \cite{Aichhorn2009,Aichhorn2011} combining the full-potential linearized augmented plain-wave (FLAPW) 
band structure method \cite{Wien2k} with the dynamical mean-field theory (DMFT) \cite{Georges1996} treatment of the on-site Coulomb 
repulsion between Fe 3$d$ states. A combination of LDA and DMFT has been applied earlier to study thermodynamic 
stability \cite{Leonov2011} and to describe magnetic properties \cite{Lichtenstein2001}  
of paramagnetic $bcc$ Fe at ambient pressure.

In Fig.~\ref{fig:band_struct} we display the LDA+DMFT $\vk$-resolved spectral functions $A(\vk,E)$ for the
three phases obtained for
the temperature of 5800~K. First, one may notice
that in $hcp$ Fe the electronic states in the vicinity of $E_F$ are sharp (their
red color indicating high value of $A(\vk,E)$), hence $\epsilon-$Fe exhibits a typical behaviour of a Fermi-liquid (FL) 
with large quasi-particle life-times in the vicinity of $E_F$. In contrast, the $bcc$ phase features 
a low-energy $e_g$ band along the $N-\Gamma-P$ path that is strongly broadened, thus 
indicating destruction of quasiparticle states. $fcc$ Fe is in an intermidiate state, with some broadening
noticable in the $e_g$ bands at $E_F$ in the vicinity of the $\Gamma$ and $W$ points.

\begin{figure}
\begin{center}
\includegraphics[width=0.75\columnwidth]{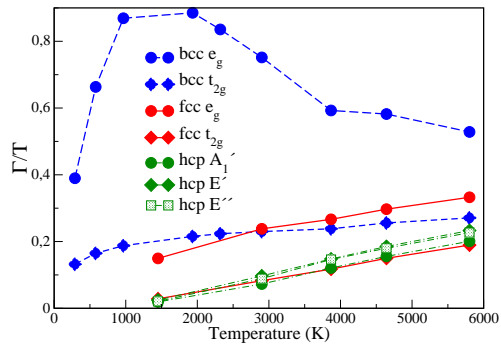}
\caption{\label{fig:gamma} (Color online)
The ratio of the inverse quasiparticle lifetime $\Gamma$ to temperature $T$ vs. $T$.
The solid red, dashed blue and dash-dotted green curves correspond to 3d states in $fcc$, $bcc$, and $hcp$ Fe,
respectively. They are split by the crystal field into $t_{2g}$ (diamonds) and $e_g$ (circles) representations in the cubic
($bcc$ and $fcc$) phases, and two doubly-degenerate ($E'$ and $E''$, shown by diamonds and squares, respectively) and 
one singlet ($A_1'$, circles) representations in the
$hcp$ phase, respectively . A non-linear behavior of $\Gamma$/T for $bcc$ Fe $e_g$ states  is clearly seen.
}
\end{center}
\end{figure}

To quantify the degree of non-Fermi-liquid (non-FL) behaviour we have evaluated the inverse quasiparticle life-time
$\Gamma=-Z\Im\left[\Sigma(i0^+)\right]$, 
where the quasiparticle residue $Z^{-1}=1-\frac{\partial \Im \Sigma(i\omega)}{\partial \omega}|_{\omega \to 0^+}$, by extrapolating 
the imaginary-frequency self-energy $\Sigma(i\omega)$ to zero. In the FL regime $\Gamma$ scales as $T^2$, hence
$\Gamma/T$ vs. $ T$ is linear. In Fig.~\ref{fig:gamma} we display the temperature evolution of $\Gamma/T$ for 
the relevant irreducible representations of the Fe 3$d$ shell in all three phases. One may see that $\Gamma/T$  in $hcp$ Fe exhibits a linear
increase typical for a FL up to temperatures expected in the inner Earth core. In contrast,  $\Gamma/T$ for the $bcc$ iron $e_g$ states
features a linear and steep rise for $T < $ 1000~K and then behaves  non-linearly, indicating a
non-coherent nature of those states at high temperatures. The $bcc$ Fe $t_{2g}$ and $fcc$ Fe $e_g$ electrons
are in an intermediate situation with some noticeable deviations from the FL behaviour.

The tendecy of $bcc$ $e_g$ states to a non-FL behaviour  has been noted before for ambient conditions 
and explained by a smaller effective bandwidth of the "localized" $e_g$ band compared to the $t_{2g}$ one \cite{Katanin2010}. 
We have evaluated the one-electron kinetic energy of the  $e_g$ and $t_{2g}$ bands as $E_b=\int_{-\infty}^{E_F} D(E) (E-C) dE$, 
where $D(E)$ is the corresponding LDA partial density of states (PDOS), $C$ is the centralweight of the band. 
Resulting $E_b$ for the $e_g$ and $t_{2g}$ bands in the $bcc$ ($fcc$) phases are equal to -1.05 (-1.01) and -1.08 (-1.20) eV, respectively. 
One may see that  the difference in kinetic energy between  the $e_g$ and $t_{2g}$ bands in $bcc$ Fe is rather small and, in fact, even smaller than the corresponding difference in the $fcc$ phase. 
Hence, it can hardly explain the observed qualitatevely distinct non-FL behaviour of the $e_g$ states in $bcc$. 
It has been pointed out \cite{Maglic1973,Irkhin1993} that a van Hove singularity in one of $bcc$ Fe $e_g$ bands leads to formation of a narrow peak in the corresponding PDOS in the vicinity of $E_F$ (see SI Fig.~S1). 
A large peak in PDOS located at $E_F$ leads to suppression of the low-energy hopping and to the corresponding enhancement of correlations, as has been recently pointed out for the case of Sr$_2$RuO$_4$ \cite{Mravlje2011}. A similar suppression is observed in the $e_g$ hybridization function in $bcc$ Fe (see SI Fig.~S2).

Having demonstrated the impact of correlation effects on the Fe electronic structure we now focus on its consequences for the Fe phase stability and magnetism. To evaluate the impact of correlation effects on the relative stability of the three phases we have computed the corresponding correction to the fixed-lattice free energy by employing a coupling-constant integration approach (see in, e.g., Ref.~\cite{Nozieres1963}). We define the free energy $F_{\lambda}=-\frac{1}{\beta}\ln Tr\left(\exp[-\beta(H_0+\lambda H_1)]\right)$  corresponding to a given value of the coupling $\lambda \in [0:1]$, where $H_0$ is the one-electron (LDA) part of the Hamiltionian, $H_1$ is the interacting part equal to the difference between the Hubbard term $H_U$ 
and the double-counting correction $E_{dc}$. The coupling constant integration leads to the following expression for the many-body correction:

\begin{equation}\label{eq:dF}
\Delta F= F-F_0 = \int_0^1 \frac{\langle \lambda H_1 \rangle_{\lambda}}{\lambda} d\lambda
\end{equation} 

In derivation of Eq.~\ref{eq:dF} we neglected the $\lambda$  dependence of the one-electron part, and, hence, 
the charge density renormalization due to many-body effects. However, we verified that the correction 
to the total energy due to the charge density self-consistency is rather small and within our error bars.

To obtain $\Delta F$ we have computed $\frac{\langle \lambda H_1 \rangle_{\lambda}}{\lambda}$ for a discret set of values of $\lambda$ ranging from 0 to 1 by performing LDA+DMFT simulations with the Coulomb interaction scaled accordingly and evaluating $\langle \lambda H_U \rangle_{\lambda}$ in accordance with the Migdal formula (we calculated $\frac{\langle \lambda H_1
  \rangle_{\lambda}}{\lambda}|_{\lambda=0}$ analytically, as in this case it is
  equal to the Hartree-Fock approximation to $\langle H_U \rangle$ computed
  with the LDA density matrix minus the double-counting correction. The
  resulting value of $\frac{\langle \lambda H_1
  \rangle_{\lambda}}{\lambda}|_{\lambda=0}$ is small, of order of 0.1 mRy).
 . 
Then we integrated  $\frac{\langle \lambda H_1 \rangle_{\lambda}}{\lambda}$  over  $\lambda$ numerically in order to obtain  $\Delta F$ and its error bar.  

\begin{figure}
\begin{center}
\includegraphics[width=0.8\columnwidth]{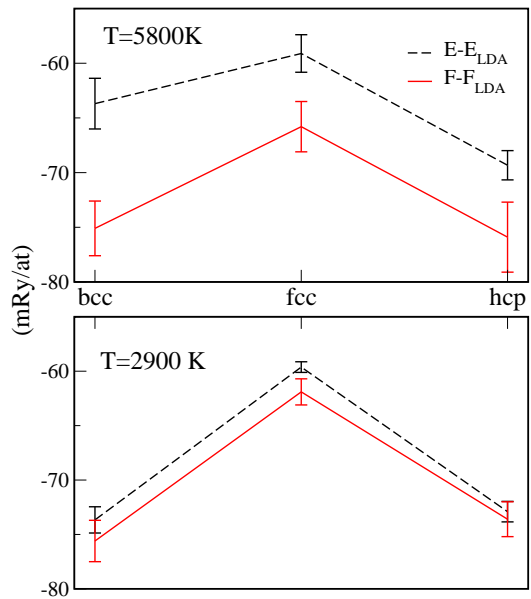}
\caption{\label{fig:free_enr} (Color online)
Many-body correction to the total (black dashed line) and free (red solid line) energy for the
three phases of Fe at the volume of 7.05 \AA/atom at T=5800~K (upper panel) and 2900~K (lower panel).
The error bars are due to the CT-QMC stochastic error.
}
\end{center}
\end{figure}

The resulting many-body correction to the electronic free energy is displayed in Fig.~\ref{fig:free_enr} for temperatures of 2900~K and 5800~K. There we also show the corresponding correction to the total energy $\Delta E=E_{DMFT}-E_{LDA}$, where $E_{DMFT}$ was computed in accordance with  Eq.~(3) of Ref.~\cite{Aichhorn2011}. Within our error bars the magnitude of  $\Delta F$ is the same for $bcc$ and $hcp$ Fe, which are suggested as stable phases of iron\cite{Vocaldo2003} and iron-based alloys \cite{Lin2002,Dubrovinsky2007} at the Earth's inner core conditions.   The magnitude of $\Delta F$ is at least several mRy smaller in the case of $fcc$ Fe, showing that the many-body correction may significantly affect relative energy differences among iron phases at the Earth core conditions.
One may also notice that the entropic contribution $T\Delta S =  \Delta E - \Delta F$ becomes much more significant at the higher temperature, and its contribution is almost twice larger in the case of the $bcc$ phase compared with two others.

\begin{figure}
\begin{center}
\includegraphics[width=0.8\columnwidth]{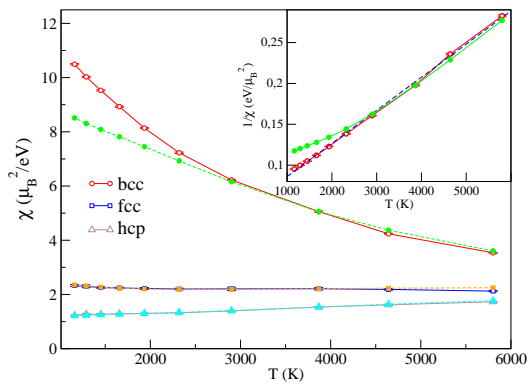}
\caption{\label{fig:suscept}
(Color online) The  uniform magnetic susceptibility in paramagnetic state
versus temperature.  The error bars are due to the CT-QMC stochastic error. The  dashed lines with corresponding filled symbols are fits to the enhanced Pauli law, see the text. Inset: the inverse uniform magnetic suscptibility of $bcc$ Fe is shown in red (empty circles), the blue dot-dashed and green (filled circles) lines are fits to the Curie-Weiss and enhanced Pauli law, respectively.}
\end{center}
\end{figure}

The application of the LDA+DMFT theory has the most important consequences for the understanding of magnetic properties of Fe at the Earth's core conditions. In Fig.~\ref{fig:suscept} we display the temperature evolution of the uniform magnetic susceptibility $\chi$ in the range of temperatures from
1100 to 5800~K.  One may notice that in the $fcc$ and $hcp$ phases the susceptibility exhibits a temperature-independent Pauli behaviour expected for a FL (a small decrease in $\chi$ at lower temperatures observed in $hcp$ Fe
  is due to a dip in its one-electron DOS in the vicinity of the Fermi level,
  see SI Fig.~S1.). In contrast,  $\chi$ of $bcc$ Fe features a clear inverse-temperature dependence and can be very well described by the Curie-Weiss (CW) law $\chi=\frac{1}{3}\frac{\mu_{eff}^2}{T+\Theta}$ with $\mu_{eff}=$2.6~$ \mu_B$ and $\Theta=$1396~$K$ (see inset in Fig.~\ref{fig:suscept}).  



One may relate the  apparent CW behaviour of the magnetic susceptibility in $bcc$ Fe to the high peak at $E_F$ present in its LDA DOS (SI Fig.~S1), which can   lead to a strongly temperature-dependent Pauli (band) susceptibility . We have computed the Stoner-enchanced Pauli susceptibilities $\chi_{st}=\chi_0/(1-I * \chi_0)$, for all three phases, where $I$ is the Stoner parameter and $\chi_0$ is the bare uniform Pauli susceptibility computed from the corresponding LDA densities of states. We fitted $I$ to reproduce the corresponding values of LDA+DMFT $\chi$ at $T=$3800~K, the resulting values of $I$ are 0.44, 0.53, and 0.54 eV in the $bcc$, $fcc$, and $hcp$ phases, respectively. Obtained $\chi_{st}$ reproduce very well the LDA+DMFT magneitic susceptibilities of $fcc$ and $hcp$ Fe, thus confirming the FL nature of these phases. In $bcc$ Fe $\chi_{st}$ describes well the CW-like behaviour of LDA+DMFT susceptibility in the range from 3000 to 5800~K.  However,  $\chi_{st}$ deviates from the LDA+DMFT susceptibility significantly at lower temperatures $T$, which are small compared to the characteristic width of the peak at $E_F$ in the LDA DOS, see Fig.~\ref{fig:suscept}.

An alternative and more interesting source for the apparent CW behaviour of the uniform susceptibility in $bcc$ Fe can be a {\it local magnetic moment} surviving in this phase up to Earth core temperatures. Existence of a local magnetic moment of the constant magnitude also provides a natural explanation for the inverse $bcc$ Fe susceptibility exhibiting the same linear temperature dependence in the whole range from 1100 to 5800~K, with no significant deviations or noticable  peculiarities (see inset in Fig~\ref{fig:suscept}).



The value of calculated uniform susceptibility in SI units at temperature T=5800~K  is equal to 1.7$*10^{-4}$, 2.0$*10^{-4}$, and 3.5$*10^{-4}$  for $hcp$, $fcc$, and $bcc$ Fe, respectively. We would like to underline that our calculated uniform magnetic susceptibilities in all three phases of Fe are sufficiently high to be important for models of the Earth core dynamics and geodynamo (see e.g. Ref.~\cite{Aubert2009}).   An inner core with a paramagnetic susceptibility in the range 10$^{-3}$ - 10$^{-4}$ SI units, and with a paramagnetic relaxation time acting slower than field changes coming from the outer core, could also attenuate short frequency fluctuations and become an important factor stabilizing the geodynamo\cite{Clement1995}, just as an electrically conducting inner core could stabilize the geodynamo because the inner core would have a magnetic diffusion constant independent of the outer core\cite{Gilder1998,Hollerback1993}. 

In conclusion, we have carried out a theoretical investigation of the role of electronic correlations in the $bcc$, $fcc$, and $hcp$ phases of Fe at the inner Earth core conditions using a fully self-consistent
LDA+DMFT approach. We have found
that the $fcc$ and $hcp$ phases remain in a Fermi-liquid state, while $bcc$ Fe features a non-Fermi-liquid behaviour.  We have evaluated a correction to the electronic free energy due to many-body effects and found
that it affects significantly the rellative free energy differences, penalizing the $fcc$ phase. 
Most interestingly, our results suggests that a local magnetic moment may exist in the $bcc$ phase at the inner core conditions and that magnetic susceptibilities in all three phases of Fe are sufficiently high to stabilize geodynamo.  Thus, new models of geodynamo as well as the core structure and elasticity should include the magnetism of the Earth's core, the effect of which so far was never considered. 

\begin{small}

\section{Methods}

In our LDA+DMFT calculations 
Wannier-like functions for the Fe-3d shell were 
constructed by projecting local
 orbitals onto a set of FLAPW Bloch states located within the energy window from -10.8 to 4 eV relative
to the Fermi level $E_F$ (details 
of the projection procedure can be found in Ref.~\cite{Aichhorn2009}). 

We then introduced
the calculated local Coulomb interaction in the density-density form acting between those Wannier orbitals. 
In order to evaluate the strength of the on-site electron repulsion on the Fe 3$d$ shell we employed the constrained 
random-phase-approximation (cRPA) method \cite{Aryasetiawan2004,Miyake2008}.
The calculated Coulomb ({\bf\it U}) and exchange ({\bf\it J}) interaction matrices are well approximated by a spherically-symmetric form  used
in the subsequent calculations, with 
the parameter $U$ (the Slater parameter $F_0$) equal to 3.15, 3.04, and 3.37 eV and the Hund's rule coupling $J$ 
equal to 0.9, 0.9 and 0.93 eV for $bcc$, $fcc$, and $hcp$ phases, respectively.  These results are in general agreement with previous calculations  of $U$ in compressed Fe \cite{Cococcioni2005}. 
We employed the around mean-field form \cite{Czyzyk1994} of the double counting correction term throughout.

The resulting many-body 
problem has been treated within the DMFT framework with the quantum impurity problem solved by the exact Continuous-time
hybridization expansion Quantum Monte-Carlo (CT-QMC) method \cite{Gull2011} 
using 5*10$^8$ CT-QMC moves with a measurement performed after each 200 moves. 
After completing the DMFT cycle we calculated the resulting density matrix in the Bloch 
states' basis, which was then used to recalculate the charge density in 
 the next iteration as described in Ref.~\cite{Aichhorn2011}. To obtain the spectral function at the real axis we employed a stochastic version 
of the Maximum Entropy method \cite{Beach2004} for analytical continuation. In order to compute the magnetic susceptibility in uniform fields we performed LDA+DMFT simulations 
with the Kohn-Sham eigenstates split by the magnetic field $H = 0.005 eV/\mu_B$ directed along the $z$ axis.

\begin{acknowledgments}
We are greaful to J. Mravlje, V. Vildosola, and A. Georges for useful
discussions.
We acknowledge the funding provided by Swedish e-science Research Centre (SeRC), Swedish Research
Centre for Advanced Functional Materials (AFM), Link\"oping Linnaeus Initiative for
Novel Functional Materials (LiLI-NFM), SRL grant 10-0026 from the
Swedish Foundation for Strategic Research (SSF), the Swedish Research Council (VR) grant 621-2011-4426 and PHD DALEN 2012 project 26228RM as well as financial support from German
Science Foundation (DFG) and German Ministry for Education and Research (BMBF). 
Calculations have been performed using the facilities of the National
Supercomputer Centre in Link\"oping (NSC) and High Performance Computing Center
North (HPC2N) at Swedish National Infrastructure for Computing (SNIC). 
\end{acknowledgments}

\paragraph{author contributions}
L.~V.~P. performed the electronic structure, susceptibility and free energy calculations, T.~M. performed the calculations of the screened Coulomb interactions.  A.~I.~A. and  L.~D. coordinated the project. All authors contributed to the results' analysis and writing of the article.

\end{small}


\begin{thebibliography}{10}

\bibitem{Lin2002}
Lin JF, Heinz DL, Campbell AJ, Devine JM, Shen G
\newblock (2002) Iron-silicon alloy in earth's core?
\newblock \emph{Science} 295:313.

\bibitem{Dubrovinsky2007}
Dubrovinsky L, {et~al.}
\newblock (2007) Body-centered cubic iron-nickel alloy in earth's core.
\newblock \emph{Science} 316:1880--1883.

\bibitem{Belonoshko2003}
Belonoshko AB, Ahuja R, Johansson B
\newblock (2003) Stability of the body-centred-cubic phase of iron in the
  earth's inner core.
\newblock \emph{Nature} 424:1032.

\bibitem{Vocaldo2003}
Vo\u{c}aldo L, {et~al.}
\newblock (2003) Possible thermal and chemical stabilization of
  body-centred-cubic iron in the earth's core.
\newblock \emph{Nature} 424:536.

\bibitem{Mikhaylushkin2007}
Mikhaylushkin AS, {et~al.}
\newblock (2007) Pure iron compressed and heated to extreme conditions.
\newblock \emph{Phys. Rev. Lett.} 99:165505.

\bibitem{Stixrude2012}
Stixrude L
\newblock (2012) Structure of iron to 1~gbar and 40\,000~k.
\newblock \emph{Phys. Rev. Lett.} 108:055505.

\bibitem{Sola2009}
Sola E, Brodholt JP, Alf\`e D
\newblock (2009) Equation of state of hexagonal closed packed iron under
  earth's core conditions from quantum monte carlo calculations.
\newblock \emph{Phys. Rev. B} 79:024107.

\bibitem{Sola2009_1}
Sola E, Alf\`e D
\newblock (2009) Melting of iron under earth's core conditions from diffusion
  monte~carlo free energy calculations.
\newblock \emph{Phys. Rev. Lett.} 103:078501.

\bibitem{Glazyrin2013}
Glazyrin K, {et~al.}
\newblock Importance of correlation effects in hcp iron revealed by a pressure-induced electronic topological transition.
\newblock \emph{Phys. Rev. Lett., in press, arXiv:1204.5130}.

\bibitem{Ono2010}
Ono S, Kikegawa T, Hirao N, Mibe K
\newblock (2010) High-pressure magnetic transition in hcp-fe.
\newblock \emph{American Mineralogist} 95:880.

\bibitem{Aichhorn2009}
Aichhorn M, {et~al.}
\newblock (2009) Dynamical mean-field theory within an augmented plane-wave
  framework: Assessing electronic correlations in the iron pnictide lafeaso.
\newblock \emph{Phys. Rev. B} 80:085101.

\bibitem{Aichhorn2011}
Aichhorn M, Pourovskii L, Georges A
\newblock (2011) Importance of electronic correlations for structural and
  magnetic properties of the iron pnictide superconductor lafeaso.
\newblock \emph{Phys. Rev. B} 84:054529.

\bibitem{Wien2k}
Blaha P, Schwarz K, Madsen G, Kvasnicka D, Luitz J
\newblock (2001) \emph{WIEN2k, An augmented Plane Wave + Local Orbitals Program
  for Calculating Crystal Properties}
\newblock (Techn. Universitat Wien, Austria, ISBN 3-9501031-1-2.).

\bibitem{Georges1996}
Georges A, Kotliar G, Krauth W, Rozenberg MJ
\newblock (1996) Dynamical mean-field theory of strongly correlated fermion
  systems and the limit of infinite dimensions.
\newblock \emph{Rev. Mod. Phys.} 68:13--125.

\bibitem{Leonov2011}
Leonov I, Poteryaev AI, Anisimov VI, Vollhardt D
\newblock (2011) Electronic correlations at the
  $\alpha$-$\gamma$ structural phase transition in
  paramagnetic iron.
\newblock \emph{Phys. Rev. Lett.} 106:106405.

\bibitem{Lichtenstein2001}
Lichtenstein AI, Katsnelson MI, Kotliar G
\newblock (2001) Finite-temperature magnetism of transition metals: An
  \textit{ab initio} dynamical mean-field theory.
\newblock \emph{Phys. Rev. Lett.} 87:067205.

\bibitem{Katanin2010}
Katanin AA, {et~al.}
\newblock (2010) Orbital-selective formation of local moments in
  $\alpha$-iron: First-principles route to an effective model.
\newblock \emph{Phys. Rev. B} 81:045117.

\bibitem{Maglic1973}
Maglic R
\newblock (1973) Van hove singularity in the iron density of states.
\newblock \emph{Phys. Rev. Lett.} 31:546--548.

\bibitem{Irkhin1993}
Irkhin VY, Katsnelson MI, Trefilov AV
\newblock (1993) On the microscopic model of fe and ni: the possible breakdown
  of the ferromagnetic fermi-liquid picture.
\newblock \emph{Journal of Physics: Condensed Matter} 5:8763.

\bibitem{Mravlje2011}
Mravlje J, {et~al.}
\newblock (2011) Coherence-incoherence crossover and the mass-renormalization
  puzzles in ${\mathrm{sr}}_{2}{\mathrm{ruo}}_{4}$.
\newblock \emph{Phys. Rev. Lett.} 106:096401.

\bibitem{Nozieres1963}
Nozi\`eres P
\newblock (1963) \emph{Theory of interacting Fermi systems}
\newblock (W. A. Benjamin, New York).

\bibitem{Aubert2009}
Aubert J, Labrosse S, Poitou C
\newblock (2009) Modelling the palaeo-evolution of the geodynamo.
\newblock \emph{Geophys. J. Int.} 169:1414--1428.

\bibitem{Clement1995}
Clement BM, Stixrude L
\newblock (1995) Inner-core anisotropy, anomalies in the time-averaged
  paleomagnetic field, and polarity transition paths.
\newblock \emph{Earth And Planetary Science Letters} 130:75--85.

\bibitem{Gilder1998}
Gilder S, Glen J
\newblock (1998) Magnetic properties of hexagonal closed-packed iron deduced
  from direct observations in a diamond anvil cell.
\newblock \emph{Science} 279:72--74.

\bibitem{Hollerback1993}
Hollerbach R, Jones C
\newblock (1993) Influence of the earth's inner core on geomagnetic
  fluctuations and reversals.
\newblock \emph{Nature} 365:541--543.

\bibitem{Aryasetiawan2004}
Aryasetiawan F, {et~al.}
\newblock (2004) Frequency-dependent local interactions and low-energy
  effective models from electronic structure calculations.
\newblock \emph{Phys. Rev. B} 70:195104.

\bibitem{Miyake2008}
Miyake T, Aryasetiawan F
\newblock (2008) Screened coulomb interaction in the maximally localized
  wannier basis.
\newblock \emph{Phys. Rev. B} 77:085122.

\bibitem{Cococcioni2005}
Cococcioni M, de~Gironcoli S
\newblock (2005) Linear response approach to the calculation of the effective
  interaction parameters in the $\mathrm{LDA}+\mathrm{U}$ method.
\newblock \emph{Phys. Rev. B} 71:035105.

\bibitem{Czyzyk1994}
Czy\ifmmode~\dot{z}\else \.{z}\fi{}yk MT, Sawatzky GA
\newblock (1994) Local-density functional and on-site correlations: The
  electronic structure of ${\mathrm{la}}_{2}$${\mathrm{cuo}}_{4}$ and
  ${\mathrm{lacuo}}_{3}$.
\newblock \emph{Phys. Rev. B} 49:14211--14228.

\bibitem{Gull2011}
Gull E, {et~al.}
\newblock (2011) Continuous-time monte~carlo methods for quantum impurity
  models.
\newblock \emph{Rev. Mod. Phys.} 83:349--404.

\bibitem{Beach2004}
{Beach} KSD
\newblock (2004) {Identifying the maximum entropy method as a special limit of
  stochastic analytic continuation}.
\newblock cond-mat/0403055.

\end{thebibliography}

\onecolumngrid
\mbox{}

\setcounter{figure}{0} 
\renewcommand{\thefigure}{S\arabic{figure}}
\newpage

\begin{center}
\Large{Electronic properties and magnetism of iron at the Earth's inner core conditions} 
\vspace{5 mm}
\large{by L.~V.~Pourovskii, T.~Miyake, S.~I.~Simak, A.~V.~Ruban, L.~Dubrovinsky, I.~A.~Abrikosov}
\vspace{10 mm}
\LARGE{Supporting information}
\end{center}



\begin{figure}[b]
\begin{center}
\includegraphics[width=0.60\columnwidth]{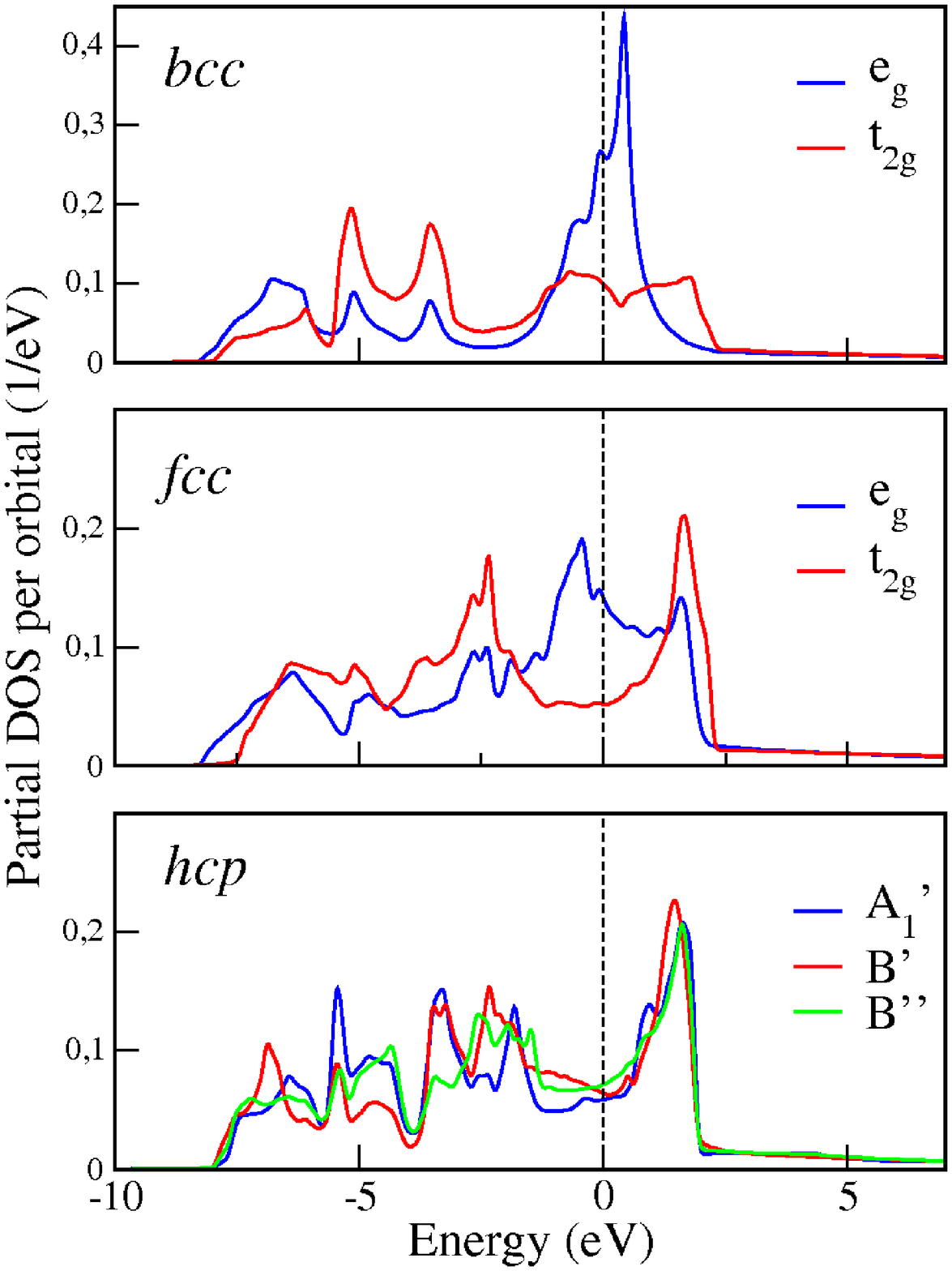}
\end{center}
\caption{\label{fig:LDA_DOS}\large
Partial LDA densities of states (PDOS) for the irreducible representations of Fe 3$d$ states 
for the three phases at the volume 7.05 \AA$^3$/atom. 
The large peak in the vicinity of $E_F$ in $bcc$ Fe $e_g$ PDOS is due to a van Hove singuliarity.}
\end{figure}

\begin{figure}
\begin{center}
\includegraphics[width=0.60\columnwidth]{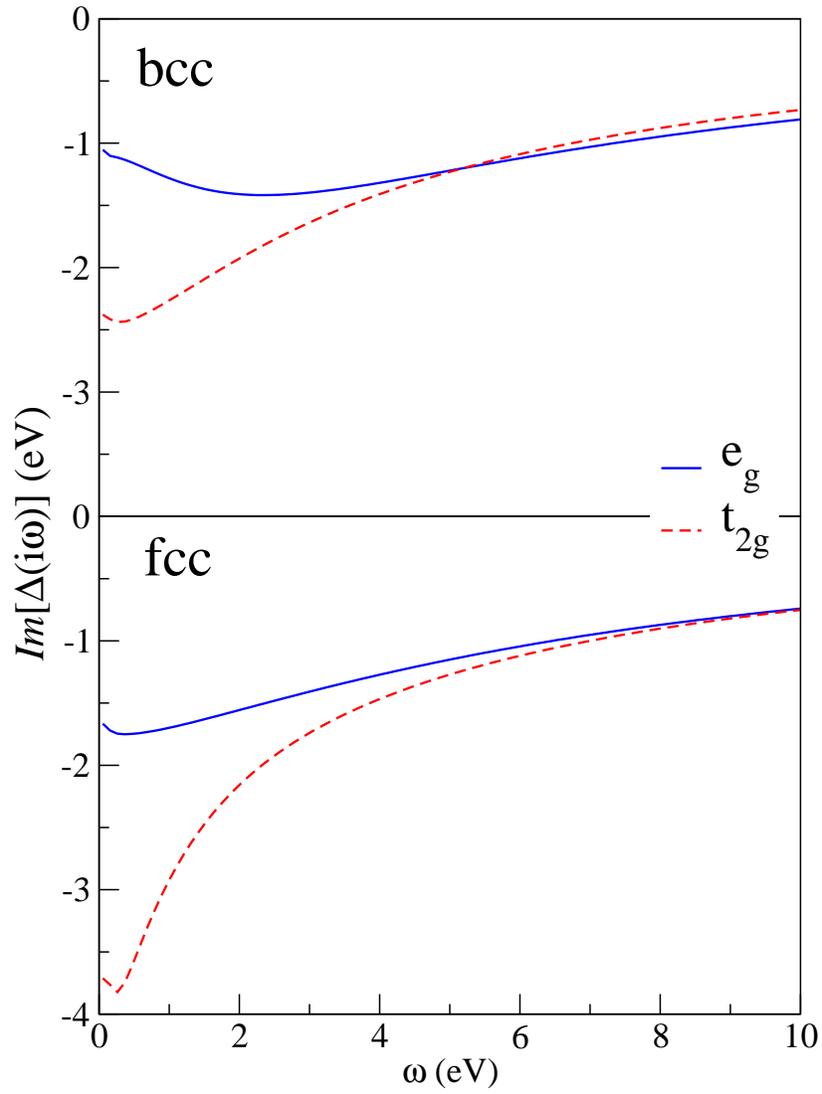}
\end{center}
\caption{\label{fig:Delta}\large
Imaginary part of the hybridization function $\Delta$ as function of imaginary frequency for the
$e_g$ and $t_{2g}$ states at the first DMFT iteration. One may clearly see a decrease in $|\Im \Delta(i\omega)|$ 
 at $\omega < 2$~eV of the $e_g$ states in $bcc$ Fe.
However, at higher energies ($\omega > 5$~eV) the
$e_g$ hybridization function decays slowly  and becomes larger than the $t_{2g}$ one. The overall
one-electron kinetic energies of $e_g$ and $t_{2g}$ states in $bcc$ Fe have similar values as explained in the main text. 
In contrast, in the $fcc$ phase $|\Delta|$ grows monotonously with decreasing $\omega$ for both $e_g$ and $t_{2g}$} 
\end{figure}

\end{document}